\documentclass[a4paper, oneside, twocolumn, notitlepage, 10pt]{extarticle_ecoc}
\usepackage{ecoc}

\usepackage[acronym]{glossaries}
\usepackage{bm}
\usepackage{siunitx}
\usepackage{upgreek}
\usepackage{nicefrac}
\usepackage{kitcolors}
\usepackage{pgfplots}
\usepackage{siunitx}
\usepackage{units}

\usepackage[disable]{todonotes}

\usetikzlibrary{fit, positioning, calc, angles, arrows.meta}
\pgfplotsset{compat=newest}

\pgfplotsset{
  discard if/.style 2 args={
    x filter/.code={
      \edef\tempa{\thisrow{#1}}
      \edef\tempb{#2}
      \ifx\tempa\tempb
      
      \fi
    }
  },
  discard if not/.style 2 args={
    x filter/.code={
      \edef\tempa{\thisrow{#1}}
      \edef\tempb{#2}
      \ifx\tempa\tempb
      \else
      
      \fi
    }
  }
}

\let\Re\relax
\let\Im\relax
\DeclareMathOperator{\Re}{Re}
\DeclareMathOperator{\Im}{Im}
\definecolor{cb-1}{HTML}{4477AA}
\definecolor{cb-2}{HTML}{EE6677}
\definecolor{cb-3}{HTML}{228833}
\definecolor{cb-4}{HTML}{CCBB44}
\definecolor{cb-5}{HTML}{66CCEE}
\definecolor{cb-6}{HTML}{AA3377}
\definecolor{cb-7}{HTML}{BBBBBB}

\pgfplotscreateplotcyclelist{cb list}{
    cb-1,cb-2,cb-3,cb-4,cb-5,cb-6,cb-7
}

\newacronym{bps}{BPS}{blind phase search}
\newacronym{rpn}{RPN}{residual phase noise}
\newacronym{awgn}{AWGN}{additive white Gaussian noise}
\newacronym{gs}{GS}{geometrical shaping}
\newacronym{qam}{QAM}{quadrature amplitude modulation}
\newacronym{snr}{SNR}{signal to noise ratio}
\newacronym{bce}{BCE}{binary cross entropy}
\newacronym{bmi}{BMI}{bitwise mutual information}
\newacronym{gcs}{GCS}{geometric constellation shaping}
\newacronym{pgcs}{pGCS}{parameterizable geometric constellation shaping}
\newacronym{gmi}{GMI}{generalized mutual information}
\newacronym{mi}{MI}{mutual information}
\newacronym{e2e}{E2E}{end-to-end}
\newacronym{cpe}{CPE}{carrier phase estimation}
\newacronym{llr}{LLR}{log-likelikood ratio}
\newacronym{nn}{NN}{neural network}
\newacronym{tx}{Tx}{transmitter}
\newacronym{rx}{Rx}{receiver}
\newacronym{fec}{FEC}{forward error correction}
\newacronym{bmd}{BMD}{bit-metric decoder}
\newacronym{ff-nn}{FF-NN}{feed-forward neural network}
\newacronym{ber}{BER}{bit error rate}
\newacronym{dsp}{DSP}{digital signal processing}

\newcommand\extrafootertext[1]{%
    \bgroup
    \renewcommand\thefootnote{\fnsymbol{footnote}}%
    \renewcommand\thempfootnote{\fnsymbol{mpfootnote}}%
    \footnotetext[0]{#1}%
    \egroup
}

\colorlet{KITColor1}{kit-blue100}
\colorlet{KITColor2}{kit-orange100}

\makeatletter
\DeclareRobustCommand{\rvdots}{%
  \vbox{
    \baselineskip4\p@\lineskiplimit\z@
    \kern-\p@
    \hbox{.}\hbox{.}\hbox{.}
  }}
\makeatother

\begin{document}
\selectlanguage{english}    %

\title{Optimization of Geometric Constellation Shaping for Wiener Phase Noise Channels with Varying Channel Parameters}%

\author{
    Andrej Rode and Laurent Schmalen
}

\maketitle                  %

\begin{strip}
 \begin{author_descr}

   Communications Engineering Lab (CEL), Karlsruhe Institute of Technology (KIT), 
   \textcolor{blue}{\uline{rode@kit.edu}}

 \end{author_descr}
\end{strip}

\setstretch{1.1}
\renewcommand\footnotemark{}
\renewcommand\footnoterule{}
\interfootnotelinepenalty=10000 %

\begin{strip}
  \begin{ecoc_abstract}
    We present a novel method to investigate the effects of varying channel parameters on geometrically shaped constellations for communication systems employing the blind phase search algorithm. We show that introduced asymmetries significantly improve performance if adapted to changing channel parameters. %
    \textcopyright2022 The Author(s)
  \end{ecoc_abstract}
\end{strip}

\section{Introduction}
\extrafootertext{This work has received funding from the European Research Council (ERC)
under the European Union's Horizon 2020 research and innovation programme (grant agreement No. 101001899). The authors acknowledge support by the state of Baden-Württemberg through bwHPC.}
\Gls*{bps} is a state-of-the-art algorithm for blind, feed-forward, carrier phase synchronization for high-rate coherent optical communications receivers. A big advantage over decision-directed, feedback-based carrier phase synchronization algorithms is the possibility for parallel and pipelined implementation~\cite{pfauHardwareEfficientCoherentDigital2009}. When a classical square \gls*{qam} constellation is used in systems employing \gls*{bps} as their carrier phase synchronization algorithm, a phase ambiguity is introduced by rotational symmetry of the constellation. %
Additionally, classical square \gls*{qam} suffers from a penalty in achievable rate and a gap to capacity, which can be overcome with geometrically or probabilistically shaped constellations~\cite{karanovEndtoEndDeepLearning2018, gumusEndtoEndLearningGeometrical2020a, cammererTrainableCommunicationSystems2020, jonesEndtoendLearningGMI2019, jovanovicEndtoendLearningConstellation2021b, arefEndtoEndLearningJoint2022a, rodeGeometricConstellationShaping2022a}. 
Therefore, we propose to apply geometric constellation shaping to improve spectral efficiency and robustness of optical communication systems employing \gls*{bps}. In previous works, \gls*{gcs} for \gls*{bps} has been either optimized on a single set of channel parameters~\cite{rodeGeometricConstellationShaping2022a}, or on a range of sets of channel parameters~\cite{jovanovicEndtoendLearningConstellation2021b}.
In %
\cite{rodeGeometricConstellationShaping2022a}, performance is only improved for channel parameters which have been used for training, or in better channel conditions. Training on a range of channel parameters results in a constellation which is robust to varying channel parameters, but may be underperforming for good channel conditions~\cite{jovanovicEndtoendLearningConstellation2021b, rodeGeometricConstellationShaping2022a}. For both approaches, changes in the position of constellation points compared to classical square \gls*{qam} cannot easily be attributed to either performance improvement of the \gls*{bps} algorithm, or robustness to channel impairments.%
Thus we propose to apply \gls*{gcs} with an additional channel condition parameter input at the mapper and demapper to investigate the effect of varying channel parameters on \gls*{gcs} for \gls*{bps}. A similar approach with parameterizable and trainable neural mapper and demapper has been shown in~\cite{cammererTrainableCommunicationSystems2020} for \gls*{awgn} channels. This will show effects of variation in different channel parameters on constellations maximizing the \gls*{bmi}\footnote{The \gls*{bmi} is often referred to as \gls*{gmi} in the optical communications community. We prefer to use the term \gls*{bmi} due to its easier resemblence with the operational meaning.}.
We will compare the performance of the parameterized \gls*{gcs} constellation with a classical square \gls*{qam} constellation and a constellation robust to changes in parameters.

\newsavebox\neuralnetwork
\sbox{\neuralnetwork}{%
		\begin{tikzpicture}[
      >=stealth,
      scale=.9,
      every node/.append style={transform shape},
      remember picture,
      ]%
      \tikzset{Source1b/.style={rectangle, draw=black, thick, minimum width=0.05cm, minimum height=1.2cm, rounded corners=0.5mm}}
      \tikzset{Source3/.style={rectangle, draw, thick, minimum width=0.6cm, minimum height=0.45cm, rounded corners=0.5mm}}

      \tikzset{OnehotNode/.style={circle, thick, draw,minimum width=0.1cm}}
      \tikzset{ReLUNode/.style={circle,thick,draw,fill=black!10!white}}
      \tikzset{MZMNode/.style={circle,thick,draw,fill=black!30!white}}
   \node (serial) at (3.6,0) {};
	  \def\offseta{4.5}
	  \def\offsetb{0.5cm}
	  \def\offsetc{15cm}
    \def\k{0}
    \def\ki{1}
        \node [OnehotNode] (S\ki1) at ($(0,0.5)+(0,\k)$) {};
        \node [OnehotNode] (S\ki2) at ($(0,-0.5)+(0,\k)$) {};
		\node at ($(0,0)+(0,\k)$) {$\rvdots$};
		\draw [->] ($(S\ki1)+(-0.4cm,0)$) -- (S\ki1);
		\draw [->] ($(S\ki2)+(-0.4cm,0)$) -- (S\ki2);

		\node [ReLUNode] (H\ki11) at ($(1,1.5)+(0,\k)$) {};
		\node [ReLUNode] (H\ki12) at ($(1,0.5)+(0,\k)$) {};
		\node [ReLUNode] (H\ki13) at ($(1,-0.5)+(0,\k)$) {};
		\node [ReLUNode] (H\ki14) at ($(1,-1.5)+(0,\k)$) {};
		\node at ($(1,0)+(0,\k)$) [anchor=center] {$\rvdots$};

		\foreach\j in {1,2,3,4}{
		  \draw [->] (S\ki1) -- (H\ki1\j); \draw [->] (S\ki2) -- (H\ki1\j);
		};

		\node [ReLUNode] (H\ki21) at ($(2,1.5)+(0,\k)$) {};
		\node [ReLUNode] (H\ki22) at ($(2,0.5)+(0,\k)$) {};
		\node [ReLUNode] (H\ki23) at ($(2,-0.5)+(0,\k)$) {};
		\node [ReLUNode] (H\ki24) at ($(2,-1.5)+(0,\k)$) {};
		\node at ($(2,0)+(0,\k)$) [anchor=center] {$\rvdots$};

		\foreach\j in {1,2,3,4}{
		  \foreach\i in {1,2,3,4}{
		    \draw [->] (H\ki1\j) -- (H\ki2\i);
		  };
		};

		\node [MZMNode] (M\ki1) at ($(3,1)+(0,\k)$) {};
		\node [MZMNode] (M\ki2) at ($(3,0)+(0,\k)$) {};
		\node [MZMNode] (M\ki3) at ($(3,-1)+(0,\k)$) {};
		\draw [<-] ($(serial.west)+(0,\k)+(0,1)$) -- (M\ki1.east);
		\draw [<-] ($(serial.west)+(0,\k)$) -- (M\ki2.east);
		\draw [<-] ($(serial.west)+(0,\k)+(0,-1)$) -- (M\ki3.east);

		\node at ($(3,-0.5)+(0,\k)$) [anchor=center] {$\rvdots$};
		\node at ($(3,+0.5)+(0,\k)$) [anchor=center] {$\rvdots$};
		\foreach\j in {1,2,3,4}{
		  \foreach\i in {1,2,3}{
		    \draw [->] (H\ki2\j) -- (M\ki\i);
		  };
		};
\end{tikzpicture}%
}

\begin{figure*}[ht!]
    \centering
\tikzset{MUL/.style={draw,circle,append after command={
      [shorten >=\pgflinewidth, shorten <=\pgflinewidth,]
      (\tikzlastnode.north west) edge (\tikzlastnode.south east)
      (\tikzlastnode.north east) edge (\tikzlastnode.south west)
    }
  }
}
\tikzset{ADD/.style={draw,circle,append after command={
      [shorten >=\pgflinewidth, shorten <=\pgflinewidth,]
      (\tikzlastnode.north) edge (\tikzlastnode.south)
      (\tikzlastnode.east) edge (\tikzlastnode.west)
    }
  }
}
\tikzstyle{box}=[rectangle,draw=black, minimum size=7mm, inner sep=2mm, font=\footnotesize, align=center]
\tikzstyle{node}=[circle, draw=black, minimum size=5mm]
\tikzstyle{connection}=[->,>=latex]
\tikzset{%
  cblock/.style    = {draw, thick, rectangle, minimum height = 3em, minimum width = 6em,rounded corners=1mm},
  lblock/.style = {draw,thick,rectangle,minimum height=10em, minimum width=2em,
    rounded corners=1mm},
  operation/.style = {draw, thick, minimum height= 1.5em, circle},
}
\def\aend{0.25}
	\resizebox{0.8\textwidth}{!}{
		\begin{tikzpicture}[font=\huge, thick]
		    \node [lblock, minimum width=12em] (embed) at (0,1) {\LARGE \begin{tabular}{c}
		         Binary  \\
		         $\downarrow$ \\
		         One-Hot
		    \end{tabular}};

        \node [operation, below right=0.5 of embed] (tx_mult) {$\cdot$};
        \node [lblock, below=1 of embed, minimum width=12em] (txnet) {\LARGE\begin{tabular}{c} Tx-NN\\\usebox{\neuralnetwork}\end{tabular}};
        \node [left=1 of txnet,yshift=2em] (noise_tx_input) {\makebox[\widthof{$\sigma_\upphi$}][l]{$\sigma_\mathrm{n}$}};
        \node [left=1 of txnet,yshift=-2em] (lw_tx_input) {$\sigma_\upphi$};
        \node [operation, ADD, right=2.5 of tx_mult] (awgn_add) {};
        \node [below=1 of awgn_add, text depth=1.5em, text height=1.5em] (awgn_param) {$n_k$};

        \node [operation, MUL, right=0.8 of awgn_add] (phase_mult) {};
        \node [below=1 of phase_mult, text depth=1.5em, text height=1.5em] (pn_param) {$\mathrm{e}^{\mathrm{j}\varphi}$};
		\node [above=1.2 of phase_mult] (pn_param_def) {$\Delta\varphi = \mathcal{N}(0,\sigma_\upphi^2$)};

        \node [coordinate,right=0.5 of phase_mult] (switch_start) {};
        \node [coordinate,right=0.5 of switch_start] (switch_cent) {};
        \node [coordinate,above=1 of switch_cent] (switch_top) {};
        \node [coordinate,below=1 of switch_cent] (switch_bottom) {};
        \pic [draw, angle radius=15, dotted, <-, >={Latex[scale=1.1]}, very thick] {angle=switch_bottom--switch_start--switch_top};
        \node [cblock, KITColor2, right=0.75 of switch_top, minimum width=10em] (diff_bps) {\LARGE diff. BPS};
        \node [cblock, right=0.75 of switch_bottom, minimum width=10em] (bps) {\LARGE BPS};
        \node [coordinate,right = 1 of bps] (end_switch_bottom) {};
        \node [coordinate,right = 0.75 of diff_bps] (end_switch_top) {};
        \node [coordinate,below = 0.75 of end_switch_top] (end_switch_cent) {};
        \node [coordinate,right = 0.5 of end_switch_cent] (end_switch_end) {};
        \pic [draw, angle radius=15, dotted, ->, >={Latex[scale=1.1]}] {angle=end_switch_top--end_switch_end--end_switch_bottom};
		\node [lblock,right=3 of end_switch_end] (complexrx) {\LARGE\begin{tabular}{c}{Rx-NN}\\\usebox{\neuralnetwork}\end{tabular}};
		\node [right=0.6 of complexrx,align=left] (sink) {$\hat{L}_{b,1}$\\ $\hat{L}_{b,2}$\\ $\vdots$\\ $\hat{L}_{b,m}$};
		\node [left=1 of embed,align=left] (source) {$b_1$\\ $b_2$\\ $\vdots$\\ $b_m$};
		\draw [-{Latex[scale=1.25]},thick] (source) -- (embed);
		\draw [-{Latex[scale=1.25]},thick] (awgn_param) -- (awgn_add);
		\draw [-{Latex[scale=1.25]},thick] (pn_param) -- (phase_mult);
		\draw [-{Latex[scale=1.25]},thick] (noise_tx_input)--([yshift=2em]txnet.west);
		\draw [-{Latex[scale=1.25]},thick] (lw_tx_input)--([yshift=-2em]txnet.west);
		\draw [-{Latex[scale=1.25]},thick] (txnet.east) -| (tx_mult);
        \draw [-{Latex[scale=1.25]},thick] (embed.east) -| (tx_mult);
        \draw [-{Latex[scale=1.25]},thick] (tx_mult) -- node[above left]{$x_k$} (awgn_add);
        \draw [-{Latex[scale=1.25]},thick] (awgn_add) -- (phase_mult);
        \draw [-{Latex[scale=1.25]},thick] (phase_mult) -- node[above]{$z_k$} (switch_start) -- (switch_top);
        \draw [{Circle[open]}-, thick] (switch_top) -- (diff_bps);
        \draw [{Circle[open]}-,thick] (switch_bottom) -- (bps);
        \draw [-{Circle[open]},thick] (bps) -- (end_switch_bottom);
        \draw [-{Circle[open]},thick] (diff_bps) -- (end_switch_top);
        \draw [-{Latex[scale=1.25]},thick] (end_switch_top) -- (end_switch_end) -- node[above,near start,name=hat_xk]{$\hat{x}_k$} node[yshift=2em,xshift=1em,name=noise_rx_input]{\makebox[\widthof{$\sigma_\upphi$}][l]{$\sigma_\mathrm{n}$}} node[yshift=-2em,xshift=1em,name=lw_rx_input]{$\sigma_\upphi$} (complexrx);
        \draw [-{Latex[scale=1.25]},thick] (noise_rx_input) -- ([yshift=2em]complexrx.west);
        \draw [-{Latex[scale=1.25]},thick] (lw_rx_input) --([yshift=-2em]complexrx.west);
        \draw [-{Latex[scale=1.25]},thick] (complexrx) -- (sink);

		    \node (rectch) [fit={($(awgn_add.north west)+(-1,3)$) ($(end_switch_end.south east)+(1.2,-3.2)$)}, draw, rounded corners=6pt, dashed, KITColor1,inner sep=0pt,rectangle] {};
		    \node at (rectch.north) [anchor=north,KITColor1,above] {Auto-encoder channel};
		\end{tikzpicture}}
  \captionof{figure}{System model of parameterizable auto-encoder with differentiable \gls*{bps}}
  \label{fig:system_model}
\end{figure*}
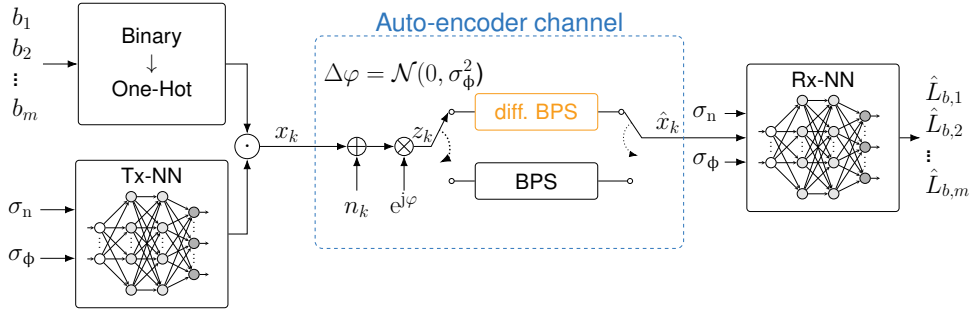

\section{Parameterizable Binary Auto-Encoder}
The system model in Fig.~\ref{fig:system_model} used for our work is an \gls*{e2e} model representing a high-rate coherent optical communication system with a transmission channel affected by \gls*{awgn} and laser phase noise. We use a %
binary auto-encoder to perform the \gls*{gcs}, which uses bit vectors of length $m$ bits per symbol as input and outputs $m$ \glspl*{llr}. At the \gls*{tx}, $m$ information bits $\bm{b}_k$ are converted to a one-hot vector. The \gls*{awgn} standard deviation $\sigma_\mathrm{n}$ and Wiener phase noise increments' standard deviation $\sigma_\upphi$ are fed as inputs to the \gls*{tx}-\gls*{nn} to generate a constellation $\mathcal{M}$ of size \mbox{$\left|\mathcal{M}\right| = 2^m$}. With the output of the \gls*{tx}-\gls*{nn} and the one-hot vector, one complex constellation point $x_k \in \mathcal{M}$ is selected. This is represented in the system model in Fig.~\ref{fig:system_model} as dot product $(\odot)$ between the one-hot vector and the constellation vector. The complex symbol vector $\bm{x} = (x_0, \ldots, x_{B-1})^\mathrm{T}$ for one batch of size $B$ is sent through the auto-encoder channel model. In our system model the auto-encoder channel is comprised of \gls*{awgn}, Wiener phase noise and a differentiable \gls*{bps}. The differentiable \gls*{bps} differs from the classical, non-differentiable, \gls*{bps} algorithm in the replacement of the $\arg\max$ operation with a differentiable approximation~\cite{rodeGeometricConstellationShaping2022a}. At the output of the (differentiable) \gls*{bps}, complex symbols $\hat{\bm{x}}$ are sent to a neural demapper with a \gls*{rx}-\gls*{nn}, which returns $m$ \glspl*{llr}. The \gls*{tx}-\gls*{nn} consists of an input dimension of \num{2} and two fully connected layers with output dimension of $2^{m+1}$ with ReLU activation functions for the input and hidden layer. %
In contrast to the system model shown in~\cite{rodeGeometricConstellationShaping2022a}, both neural mapper and demapper have additional inputs for $\sigma_\mathrm{n}$ and $\sigma_\Upphi$, which allows the \gls*{gcs} to be optimized across a range of channel conditions. At the \acrfull*{tx}, an optimized constellation for a particular channel condition can be obtained by evaluating the output of the \gls*{tx}-\gls*{nn}. We implemented our system for simulation and validation in the PyTorch framework~\cite{NEURIPS2019_bdbca288}. 

\section{Parameterized \Acrshort*{gcs} Setup}
To learn a geometrically shaped constellation which is optimized on a range of channel conditions, defined by a pair of parameters $\sigma_\mathrm{n}$ and $\sigma_\phi$, we take the following approach: For every batch in a training epoch, a new $\sigma_\mathrm{n}$ and $\sigma_\Upphi$ is sampled uniformly in $\left[\sigma_{\mathrm{n},\min},\sigma_{\mathrm{n},\max}\right]$ and $\left[\sigma_{\upphi,\min},\sigma_{\upphi,\max}\right]$. The \gls*{bps} algorithm is updated accordingly with the transmit constellation obtained with the new parameters. In this work, we selected $\sigma_\mathrm{n}$ such that the \gls*{snr} is between \SI{14}{dB} and \SI{25}{dB} and $\sigma_\Upphi$ is selected such that the laser linewidth is between \SI{50}{kHz} and \SI{600}{kHz} for a symbol rate of \SI{32}{GBaud}. \gls*{bps} has been configured with 60 test angles and a window size of \num{128}. Training has been performed for 1000 epochs with a linearly increasing batchsize from 1000 to 10000 samples. Similar to~\cite{rodeGeometricConstellationShaping2022a}, the temperature parameter of the differentiable \gls*{bps} is decreased from \num{1.0} to \num{0.001} during training to approximate the real \gls*{bps} more closely at the end of the training. The \gls*{bce} is used as loss function and the Adam algorithm~\cite{kingmaAdamMethodStochastic2015} is used for the optimization. With the binary auto-encoder, both geometric shaping and bit labelling are optimized simultaneously. For simplicity, bit labels are omitted in the constellation plots.

\begin{figure}[t!]
  \centering
  \begin{tikzpicture}%
  \pgfplotsset{colormap={kit-cm}{color={KITblue} color={KITorange}}}
    \begin{axis}[
      xlabel={$\Re\left\{x_k\right\}$},
      ylabel={$\Im\left\{x_k\right\}$},
      height=0.69\columnwidth,
      width=0.69\columnwidth,
      xmin=-2,
      xmax=2,
      ymin=-2,
      ymax=2,
      grid=both,
      colormap name=kit-cm,
      colorbar, %
      label style={font=\footnotesize},
      tick label style={font=\footnotesize},  
      colorbar style={
        ytick={50,325,600},
        ylabel={\footnotesize Laser linewidth in \si{kHz}},
      },
      scatter/use mapped color={
               mapped color!80,
           },
      ]
      \foreach\lw in{50.00,111.11,172.22,233.33,294.44,355.56,416.67,477.78,538.89,600.00}{
      \addplot[
      scatter,
      only marks,
      mark=*,
      mark size=1.2pt,
      point meta=\lw,
      ]
      table[col sep=tab]
      {data/constellation_\lw_18.txt};
      }
    \end{axis}
     \end{tikzpicture}
    \caption{Shaped constellation at \gls*{snr} of \SI{18}{dB} and varying laser linewidth}
    \label{fig:const_var_lw}
\end{figure}
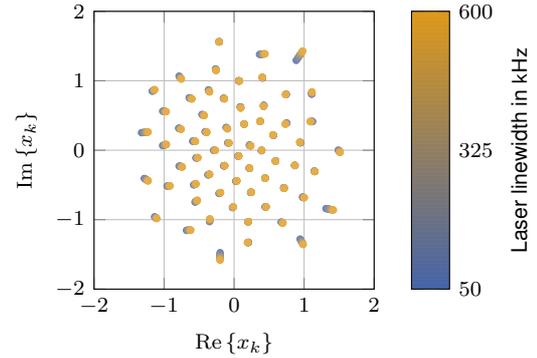

\begin{figure}[t!]
  \centering
  \begin{tikzpicture}%
  \pgfplotsset{colormap={kit-cm}{color={KITblue} color={KITorange}}}
    \begin{axis}[
      xlabel={$\Re\left\{x_k\right\}$},
      ylabel={$\Im\left\{x_k\right\}$},
      height=0.69\columnwidth,
      width=0.69\columnwidth,
      xmin=-2,
      xmax=2,
      ymin=-2,
      ymax=2,
      grid=both,
      colormap name=kit-cm,
      colorbar,%
      label style={font=\footnotesize},
      tick label style={font=\footnotesize},        
      colorbar style={
        ylabel={\footnotesize \gls*{snr} in \si{dB}},
        ytick={14, 19, 25},
      },
      scatter/use mapped color={
               mapped color!80,
           },
      ]
      \foreach\snr in{14,15,16,17,18,19,20,21,22,23,24,25}{
      \addplot[
      scatter,
      only marks,
      mark=*,
      mark size=1.2pt,
      point meta=\snr,
      ]
      table[col sep=tab]
      {data/constellation_100000.00_\snr.txt};
      }
    \end{axis}
     \end{tikzpicture}
    \caption{Shaped constellation at a laser line width of \SI{100}{kHz} and varying \gls*{snr}}
    \label{fig:constellation_var_snr}
\end{figure}
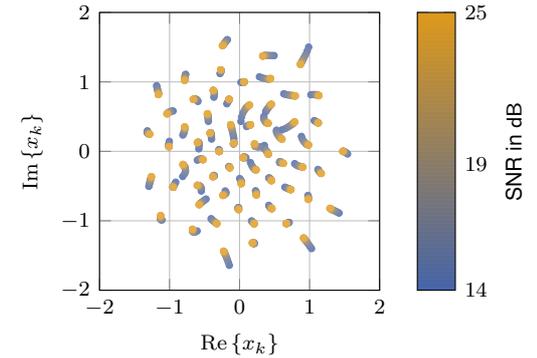

\section{Effects of Varying Channel Parameters}
To investigate the separate effect of varying \gls*{awgn} and Wiener phase noise on the shaped constellation for $m = 6$ bits per symbol, we plot some transmit constellations in Fig.~\ref{fig:const_var_lw} and Fig.~\ref{fig:constellation_var_snr}. In Fig.~\ref{fig:const_var_lw}, \gls*{awgn} is fixed to yield an \gls*{snr} of \SI{18}{dB} and the linewidth is varied between \SI{50}{kHz}, which is shown in blue, and \SI{600}{kHz}, which is shown in orange. It can be observed that most points in the transmit constellation deviate insignificantly for varying linewidth. The biggest change is observed for a single constellation point in the top right of the constellation, which is shifted outwards for a larger laser linewidth. A smaller but notable shift outwards can be observed for three constellation points in the bottom right. This observation can be explained with improved performance of the \gls*{bps} algorithm if a constellation point is separated by a larger distance radially. %
In Fig.~\ref{fig:constellation_var_snr}, we display transmit constellations for a fixed laser linewidth of \SI{100}{kHz} and the \gls*{snr} varying between \SI{14}{dB}, in blue color, and \SI{25}{dB}, in orange color. For low \glspl*{snr}, the constellation points are clustered in groups, which shows the effect of reducing the distance between constellation points differing in a single bit and increasing the distance between other constellation points. Similarly to the effects of varying linewidths in Fig.~\ref{fig:const_var_lw}, the same constellation points are moved outwards for low \gls*{snr}, which, again, points to improved \gls*{bps} performance  due to less ambiguities of the constellation impaired by phase rotations.

\begin{figure}[t!]
\begin{tikzpicture}
\pgfplotsset{
rpn/.style ={KITColor2,dashed,mark=*},
bps/.style ={KITColor1,solid,mark=square*},
}
\pgfplotsset{/pgfplots/half and half legend/.style={
        /pgfplots/legend image code/.code={%
            \draw[##1,/tikz/.cd,solid]
            (0cm,0cm) -- (0.3cm,0cm);
            \draw[##1,/tikz/.cd,dotted]
            (0.3cm,0cm) -- (0.6cm,0cm);
            \draw[##1,/tikz/.cd,dashed]
            (0.6cm,0cm) -- (0.9cm,0cm);
            },},              
}
\pgfplotsset{
    legend image with text/.style={
        legend image code/.code={%
            \node[anchor=center] at (0.3cm,0cm) {#1};
        }
    },
}
  \begin{axis}[
    xlabel={Linewidth (\si{kHz})},
    ylabel={BMI (\unitfrac{bit}{symbol})},
    ymin=4,
    ymax=6,
    xmin=50,
    xmax=600,
    grid=both,
    xtick={50,200,400,600},
    xticklabels={$50$,$200$,$400$,$600$},
    label style={font=\small},
    tick label style={font=\small},           
    width=0.75\columnwidth,
    height=0.86\columnwidth,
    mark size=1pt,
    legend style={at={(axis cs:620,4)},anchor=south west,nodes={transform shape},font=\small},
    y tick label style={xshift=-0.5ex, /pgf/number format/fixed zerofill, /pgf/number format/precision=1},
    label style={font=\small},
    legend cell align={left},
    cycle multi list={
            cb list\nextlist
            {thick, solid, mark=*},{thick, dotted,mark options={solid}, mark=x},{thick, dashed, mark options={solid}, mark=o}\nextlist
            }
    ]
    
    \addlegendimage{thick, solid, mark=*, black} \addlegendentry{perf.}
    \addlegendimage{thick, dotted, mark options={solid}, mark=x, black} \addlegendentry{over}
    \addlegendimage{thick, dashed, mark options={solid}, mark=o, black} \addlegendentry{under}

    \foreach\s in{14.00,15.00,16.00,17.00,18.00,20.00,25.00}{
    \addplot+[
    ]
    table[discard if not={snr}{\s},x expr={\thisrow{linewidth}/1000}, y=mean, col sep=space, 
    ]{data/nn_var.txt};
    \addplot+[
    ]
    table[discard if not={snr}{\s},x expr={\thisrow{linewidth}/1000}, y=mean, col sep=space, 
    ]{data/nn_var_over.txt};
    \addplot+[
    ]
    table[discard if not={snr}{\s},x expr={\thisrow{linewidth}/1000}, y=mean, col sep=space, 
    ]{data/nn_var_under.txt};
    }
  \end{axis}
  \begin{axis}[
    hide axis,
    ymin=4,
    ymax=6,
    xmin=50,
    xmax=600,
    grid=both,
    width=0.75\columnwidth,
    height=0.86\columnwidth,
    cycle multi list={cb list},
    reverse legend,
    legend style={at={(axis cs:620,6)},anchor=north west,nodes={transform shape},font=\small},
  ]
 
     \foreach\s in{14.00,15.00,16.00,17.00,18.00,20.00,25.00}{
        \addplot+[no marks, very thick] coordinates {(100,4)};
        \addlegendentryexpanded{\s}
     }
     \addlegendimage{empty legend}\addlegendentry{SNR}

  \end{axis}

\end{tikzpicture}
    \caption{Validation results of parameterizable \Acrshort*{gcs} for different \Acrshortpl*{snr}}
    \label{fig:val_param_gcs}
\end{figure}
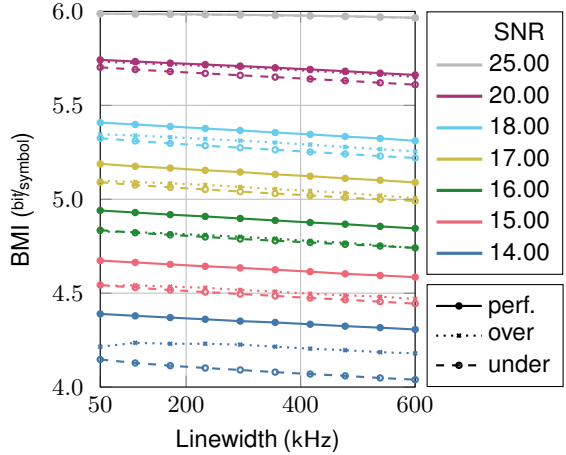

\section{Performance With Parameter Misestimation}
In Fig.~\ref{fig:val_param_gcs}, we show validation results in terms of \gls*{bmi} for \gls*{pgcs} constellations. The results are shown in the plot for a range of \glspl*{snr} and across a range of laser linewidths. The results displayed in solid lines are obtained with matching parameter inputs to the \gls*{tx}-\gls*{nn} and \gls*{rx}-\gls*{nn}. This corresponds to transmitter and receiver with perfect channel knowledge. For results with dotted lines, the \gls*{snr} is overestimated by \SI{2}{dB} at the receiver and transmitter. %
Results in dashed lines are performed with the autoencoder system underestimating the \gls*{snr} by \SI{2}{dB}. %
For high \glspl*{snr}, estimation errors in the \gls*{snr} do not incur a performance penalty with the transmitter and matched receiver system. 
Underestimating the \gls*{snr} always leads to a better \gls*{bmi} than overestimating.

\section{Performance comparison}
\begin{figure}[t!]
\begin{tikzpicture}
\pgfplotsset{
rpn/.style ={KITColor2,dashed,mark=*},
bps/.style ={KITColor1,solid,mark=square*},
}
\pgfplotsset{/pgfplots/half and half legend/.style={
        /pgfplots/legend image code/.code={%
            \draw[##1,/tikz/.cd,solid]
            (0cm,0cm) -- (0.3cm,0cm);
            \draw[##1,/tikz/.cd,dotted]
            (0.3cm,0cm) -- (0.6cm,0cm);
            \draw[##1,/tikz/.cd,dashed]
            (0.6cm,0cm) -- (0.9cm,0cm);
            },},              
}
\pgfplotsset{
    legend image with text/.style={
        legend image code/.code={%
            \node[anchor=center] at (0.3cm,0cm) {#1};
        }
    },
}
  \begin{axis}[
    xlabel={Linewidth (\si{kHz})},
    ylabel={BMI (\unitfrac{bit}{symbol})},
    ymin=4,
    ymax=6,
    xmin=50,
    xmax=600,
    grid=both,
    xtick={50,200,400,600},
    xticklabels={$50$,$200$,$400$,$600$},
    label style={font=\small},
    tick label style={font=\small}, 
    y tick label style={xshift=-0.5ex, /pgf/number format/fixed zerofill, /pgf/number format/precision=1},
    width=0.75\columnwidth,
    height=0.86\columnwidth,
    mark size=1pt,
    legend style={at={(axis cs:620,4)},anchor=south west,nodes={transform shape},font=\small},
    label style={font=\small},
    legend cell align={left},
    cycle multi list={
            cb list\nextlist
            {thick, solid, mark=*},{thick, dotted,mark options={solid}, mark=x},{thick, dashed, mark options={solid}, mark=o}\nextlist
            }
    ]
    
    \addlegendimage{thick, solid, mark=*, black} \addlegendentry{pGCS}
    \addlegendimage{thick, dotted, mark options={solid}, mark=x, black} \addlegendentry{QAM}
    \addlegendimage{thick, dashed, mark options={solid}, mark=o, black} \addlegendentry{robust}

    \foreach\s in{14.00,15.00,16.00,17.00,18.00,20.00,25.00}{
    \addplot+[
    ]
    table[discard if not={snr}{\s},x expr={\thisrow{linewidth}/1000}, y=mean, col sep=space, 
    ]{data/nn_var.txt};
    \addplot+[
    ]
    table[discard if not={snr}{\s},x expr={\thisrow{linewidth}/1000}, y=mean, col sep=space, 
    ]{data/qam_nn_var.txt};
    \addplot+[
    ]
    table[discard if not={snr}{\s},x expr={\thisrow{linewidth}/1000}, y=mean, col sep=space, 
    ]{data/nn_robust.txt};
    }
  \end{axis}
  \begin{axis}[
    hide axis,
    ymin=4,
    ymax=6,
    xmin=50,
    xmax=600,
    grid=both,
    xtick={50,200,400,600},
    xticklabels={$50$,$200$,$400$,$600$},
    width=0.75\columnwidth,
    height=0.86\columnwidth,
    label style={font=\small},
    tick label style={font=\small},
    mark size=1pt,
    cycle multi list={cb list},
    reverse legend,
    legend style={at={(axis cs:620,6)},anchor=north west,nodes={transform shape},font=\small,minimum width=2.8em},
    legend cell align={left},
  ]
 
     \foreach\s in{14.00,15.00,16.00,17.00,18.00,20.00,25.00}{
        \addplot+[no marks, very thick] coordinates {(100,4)};
        \addlegendentryexpanded{\s}
     }
     \addlegendimage{empty legend}\addlegendentry{SNR}

  \end{axis}

\end{tikzpicture}
    \caption{Performance comparison of parametrizable \Acrshort*{gcs} with square QAM and robust constellation from~\cite{jovanovicEndtoendLearningConstellation2021b}}
    \label{fig:val_compare}
\end{figure}
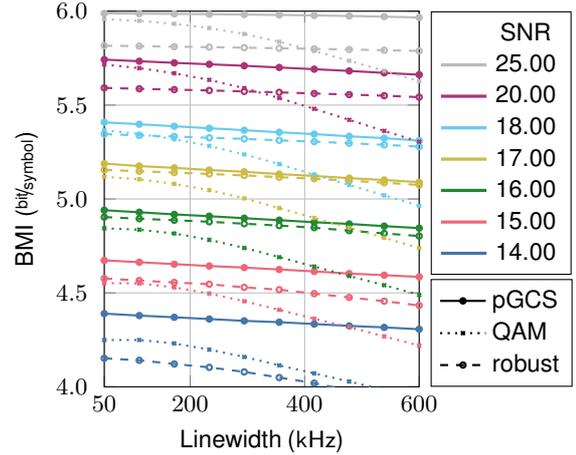

In Fig.~\ref{fig:val_compare}, the validation results of the \gls*{pgcs} constellation, a Gray-mapped square \gls*{qam} and a constellation robust to variance in the channel parameters~\cite{jovanovicEndtoendLearningConstellation2021b} are compared in terms of the \gls*{bmi}. The Gray-mapped square \gls*{qam} constellation, which is combined with a matched parameterized neural demapper, and the robust constellation are trained on the same range of channel parameters as the \gls*{pgcs} constellation, but without a parameterized channel condition input. The \gls*{pgcs} constellation outperforms the reference constellations for all channel conditions. The robust constellation matches the performance of the \gls*{pgcs} constellation very closely for \SI{17}{dB} and \SI{18}{dB}, but shows a significantly worse performance for higher and lower \gls*{snr}. The robust constellation shows a good performance for increasing laser linewidth, with only an insignificant drop in performance compared to the \gls*{pgcs} constellation. The performance of the Gray-mapped \gls*{qam} constellation is close to the performance of the \gls*{pgcs} constellation for high \gls*{snr} and low laser linewidth. For increasing laser linewidths, a substantial drop in performance can be observed. 

\section{Conclusions}

In this work, we have shown the effect of changes in channel parameters for \gls*{gcs} constellations in the presence of the \gls*{bps} algorithm for carrier phase recovery. We have introduced the novel \gls*{pgcs} constellation. For higher laser linewidths and \gls*{awgn}, the introduced asymmetry of a small number of constellation points contributes significantly to performance improvement over static reference constellations.

\printbibliography

\vspace{-4mm}

\end{document}